\pdfoutput=1
\documentclass[PAPER, atlasdraft=false, REPORT=false, UKenglish, cernpreprint=true, texmf, orcidlogo]{atlasdoc}
 
\usepackage[abbrevref=false]{atlaspackage}
\usepackage{atlasbiblatex}
 
\usepackage[hepparticle,hepprocess]{atlasphysics}
\usepackage[nice]{nicefrac}

\graphicspath{{logos/}{figures/}}
 
\usepackage{ANA-EXOT-2023-01-PAPER-defs}
\usepackage{makecell}

\AtlasTitle{Combination of searches for Higgs boson decays into a photon and a massless dark photon using $pp$ collisions at $\sqrt{s} = \SI{13}{\ensuremath{\text{\TeV}}}$ with the ATLAS detector}

\AtlasAbstract{
A combination of searches for Higgs boson decays into a visible photon and a massless dark photon ($H\to\gamma\gamma_{\text{d}}$) is presented using \SI{139}{\ifb} of proton--proton collision data at a centre-of-mass energy of $\sqrt{s} = 13$~\TeV recorded by the ATLAS detector at the Large Hadron Collider.
The observed (expected) 95\% confidence level upper limit on the Standard Model Higgs boson decay branching ratio is determined to be $\mathcal{B}(H\to\gamma\gamma_{\text{d}}) < 1.3\,\%~(1.5)\,\%$.
The search is also sensitive to higher-mass Higgs bosons decaying into the same final state.
The observed (expected) 95\% confidence level limit on the \xs times branching ratio ranges from \bsmbrlimitobslomass (\bsmbrlimitexplomass) for $\mH = \SI{400}{\GeV}$ to \bsmbrlimitobshimass (\bsmbrlimitexphimass) for $\mH = \SI{3}{\TeV}$.
Results are also interpreted in the context of a minimal simplified model.
}

\AtlasRefCode{EXOT-2023-01}
 
\PreprintIdNumber{CERN-EP-2024-152}

\AtlasJournal{JHEP}
\AtlasJournalRef{\JHEP 08 (2024) 153}
\AtlasDOI{10.1007/JHEP08(2024)153}

\hypersetup{pdftitle={ATLAS document},pdfauthor={The ATLAS Collaboration}}
 
\begin{document}
\sisetup{range-phrase=--}
 
\maketitle
 
\tableofcontents

 
\section{Introduction}
\begin{figure}[ht]
\centering
\subfloat[]{\includegraphics[width=.3\textwidth]{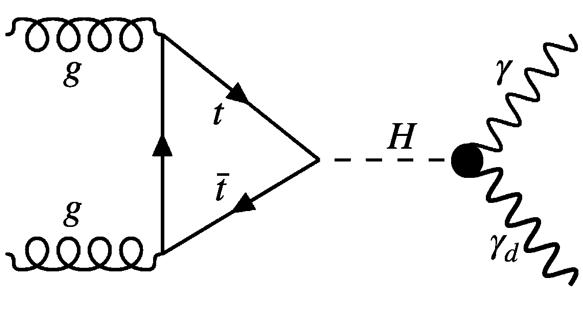}} \hspace{0.2cm}
\subfloat[]{\includegraphics[width=.3\textwidth]{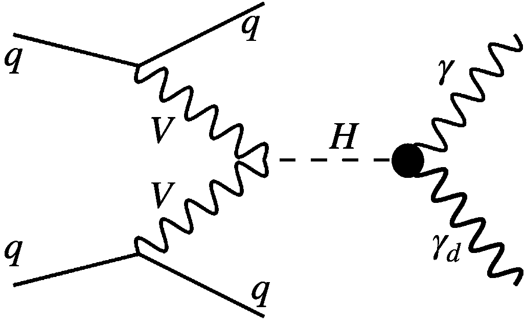}} \\
\subfloat[]{\includegraphics[width=.3\textwidth]{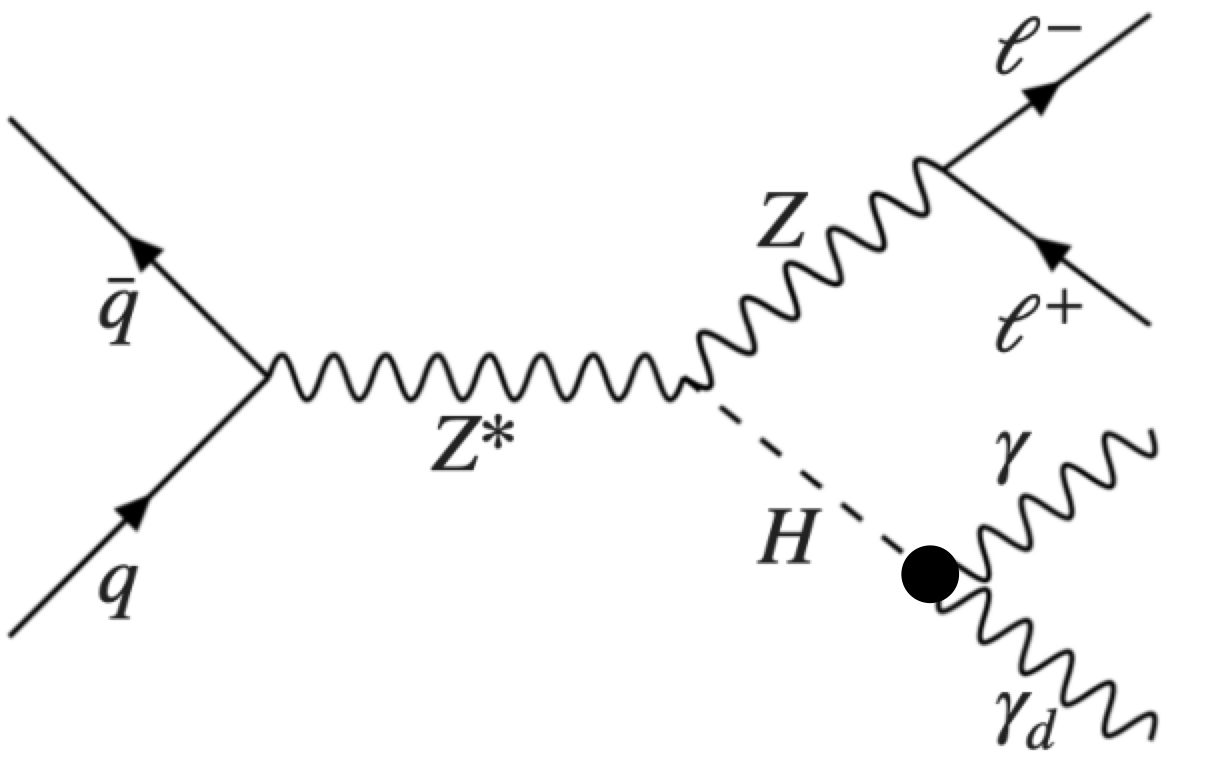}}
\subfloat[]{\includegraphics[width=.3\textwidth]{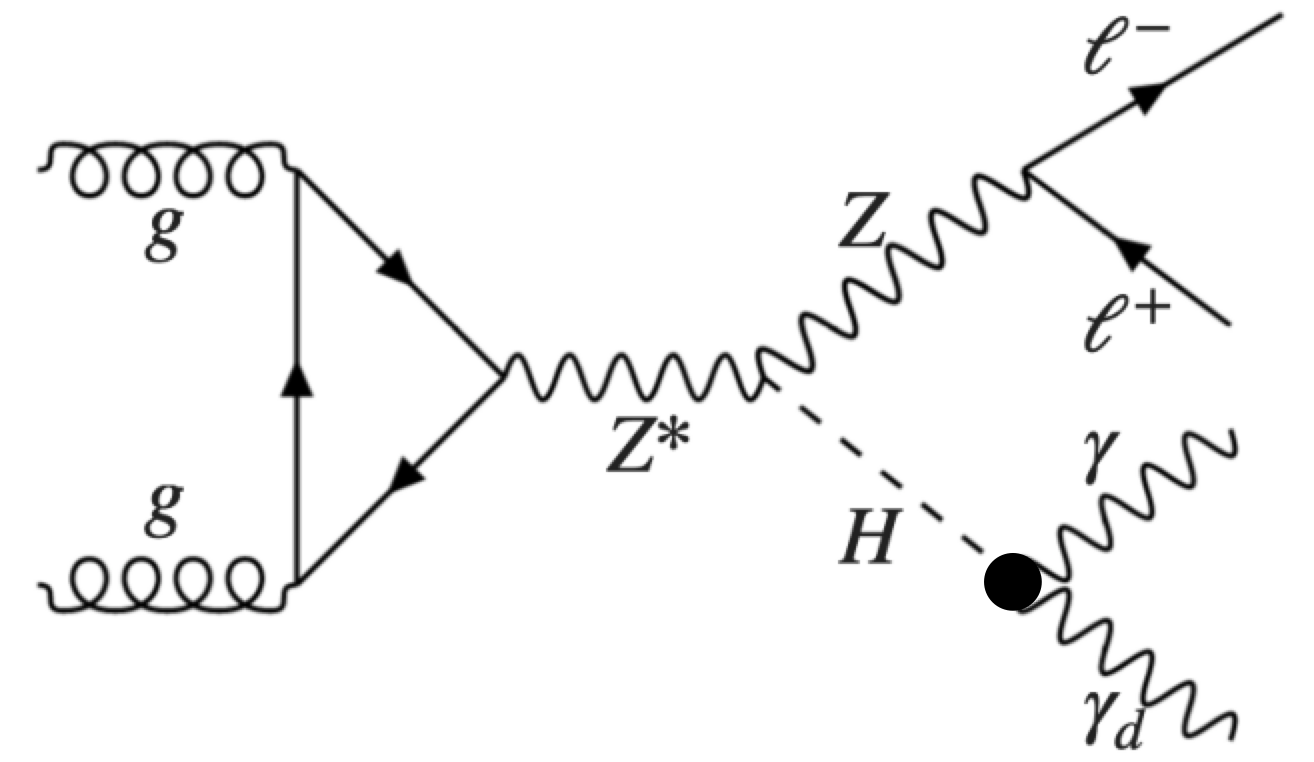}}
\subfloat[]{\includegraphics[width=.3\textwidth]{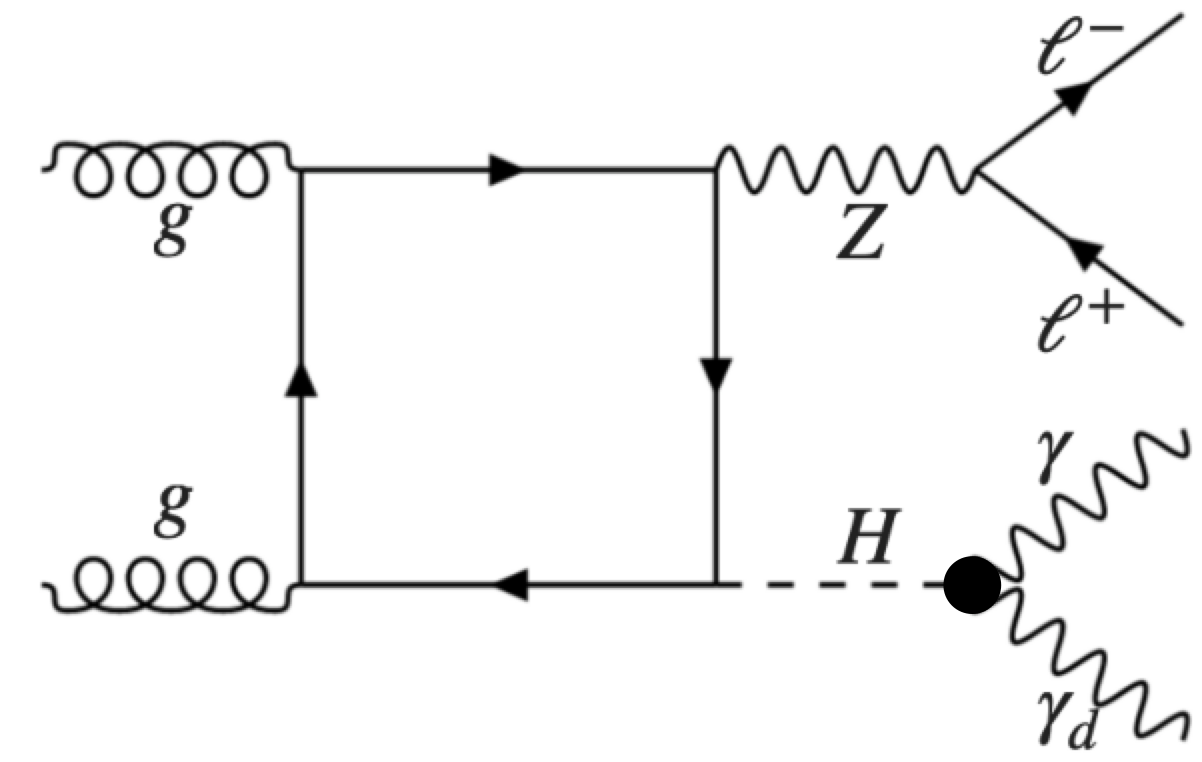}}
\caption{
Leading-order Feynman diagrams for Higgs boson production decaying into a visible photon and a massless dark photon.
The Higgs boson is produced via different processes: (a) ggF, (b) VBF, (c) $q\bar{q}\to ZH$ and (d) and (e) $gg\to ZH$.
}
\label{fig:feynman}
\end{figure}

The existence of dark matter is supported by a series of astrophysical observations~\cite{corbelli1999,Rubin:1980,Begemann:1991,Hinshaw2012,Akrami:2018vks,Trimble1987,Bertone2005,Feng2010}, but its unknown fundamental nature remains one of the central questions in particle physics.
The discovery of a Higgs boson (\PH)~\cite{HIGG-2012-27,CMS-HIG-12-028} with a mass of \SI{125}{\GeV} at the Large Hadron Collider (LHC)~\cite{Evans:2008zzb} has initiated an extensive effort aimed at investigating its possible connection with physics beyond the Standard Model (BSM).
In particular, an upper limit of the order of $\SI{10}{\%}$ on the undetected Higgs boson decay branching ratio~\cite{HIGG-2021-23,CMS-HIG-22-001} motivates searches for elusive BSM dark sector particles coupled to the Higgs boson.
One example of such particles is an undetectable, massless dark photon ($\yd$), which acts as the force carrier of an extra U(1)$_\text{d}$ gauge group of the dark sector.
Dark photons may introduce dark matter self-interactions that can potentially solve the small-scale structure formation problem~\cite{Fan:2013} and the PAMELA-Fermi-AMS2 anomaly~\cite{Arkani:2009}.
They may also play a role in enhancing the light dark matter annihilation rate to reach the required phenomenological threshold, thereby making asymmetric dark matter scenarios phenomenologically viable~\cite{Zurek:2014}.
A potential approach to search for this particle is through the Higgs boson decaying into a visible photon and a massless dark photon (\hyyd)~\cite{Gabrielli:2014oya,Biswas:2016jsh,Gabrielli_Raidal:2014,Kelly:2019,Jaeckel:2013,Beauchesne2023a,Beauchesne2023b}.
This scenario can involve either the Standard Model (SM) Higgs boson (\smhiggs) or additional high-mass Higgs bosons predicted in some BSM theories (\bsmhiggs).
As the produced \yd is invisible to the detector, its signature is characterised by missing transverse momentum, whose magnitude is denoted \met.
The same behaviour is also exhibited by ultra-light dark photons with masses below the threshold to decay into SM particles or by the lightest supersymmetry particles in a cascaded Higgs-boson decay~\cite{1997PhLB..407..243D,2012JHEP...10..016P}.
In this paper, the Higgs boson is assumed to be produced via three different processes, including gluon--gluon fusion (ggF), vector-boson fusion (VBF), and in association with a $Z$ boson ($ZH$), with the SM-like production cross-sections~\cite{deFlorian:2016spz}.
Their leading-order Feynman diagrams, including the Higgs boson decay, are shown in \cref{fig:feynman}.
 
The CMS experiment reported searches for such Higgs boson decays exploiting the $ZH$ production mechanism, with an integrated luminosity of \SI{137}{\ifb}~\cite{CMS-EXO-19-007}, or via VBF production with an integrated luminosity of \SI{130}{\ifb}~\cite{CMS-EXO-20-005}, where a combination of the two searches is also performed.
Using the \SI{8}{\TeV} collision data, the ATLAS and CMS Collaborations also reported various searches targeting the same $\gamma+\met$ final states~\cite{EXOT-2014-06,CMS-EXO-12-047,CMS-HIG-14-025}.
 
This paper presents a combined search for \hyyd produced in three distinct final-state signatures: \channelnameVBF (VBF channel\footnote{
In this paper, the term `channel' is used to refer to the final-state signature that defines the event selections.
The term `process' refers to the production mode of the Higgs boson.
Events from multiple processes can contribute to a given channel and they are all treated as signal events.
}), \channelnameZHinitialtime ($ZH$ channel) and \channelnameggF (ggF channel).
The searches under consideration use the full LHC Run~2 data sample corresponding to an integrated luminosity of \SI{139}{\ifb}~\cite{DAPR-2021-01, LUCID2} of proton--proton (\pp) collision data recorded with the ATLAS detector at a centre-of-mass energy of $\sqrt{s} = \SI{13}{\TeV}$.
The results from the VBF channel are derived based on the analysis described in Ref.~\cite{EXOT-2021-17}, with an extension to higher-mass Higgs bosons and the inclusion of the ggF process contributions.
The results for the $ZH$ channel are taken from Ref.~\cite{HDBS-2019-13}.
The results for the ggF channel are based on the reinterpretation of the mono-$\gamma$ results in Ref.~\cite{EXOT-2018-63} using the RECAST technique~\cite{2011JHEP...04..038C}.
Using the available results in the individual publications, the VBF and $ZH$ channels are combined under the SM Higgs boson assumption, and the VBF and ggF channels are combined under the additional high-mass Higgs boson hypothesis in the narrow width approximation.
The range of the \bsmhiggs masses probed starts at \SI{400}{\GeV}, as the sensitivity of the ggF channel decreases for lower masses, and goes up to \SI{3}{\TeV}.
 
The ATLAS detector~\cite{PERF-2007-01} is a multipurpose particle detector with cylindrical geometry.
It consists of an inner tracking detector surrounded by a thin superconducting solenoid providing a \qty{2}{\tesla}  axial magnetic field, electromagnetic and hadronic sampling calorimeters, and a muon spectrometer with three toroidal superconducting magnets, providing a near \(4\pi\) coverage in solid angle.\footnote{
ATLAS uses a right-handed coordinate system with its origin at the nominal interaction point (IP) in the centre of the detector and the \(z\)-axis along the beam pipe.
The \(x\)-axis points from the IP to the centre of the LHC ring, and the \(y\)-axis points upwards.
Polar coordinates \((r,\phi)\) are used in the transverse plane, \(\phi\) being the azimuthal angle around the \(z\)-axis.
The pseudorapidity is defined in terms of the polar angle \(\theta\) as \(\eta = -\ln \tan(\theta/2)\) and is equal to the rapidity $ y = \frac{1}{2} \ln \left( \frac{E + p_z c}{E - p_z c} \right) $ in the relativistic limit.
Angular distance is measured in units of \(\Delta R \equiv \sqrt{(\Delta y)^{2} + (\Delta\phi)^{2}}\).
}
An extensive software suite~\cite{SOFT-2022-02} is used in data simulation, in the reconstruction and analysis of real and simulated data, in detector operations, and in the trigger~\cite{TRIG-2016-01} and data acquisition systems of the experiment.

Monte Carlo (MC) samples of simulated signal events for the ggF, VBF, and $ZH$ via both the $q\bar{q} \to ZH$ and $gg \to ZH$ processes were generated at next-to-next-to-leading order (\NNLO) for ggF, next-to-leading order (\NLO)  for VBF and $q\bar{q} \to ZH$ or leading order (\LO) for $gg \to ZH$ in quantum chromodynamics (QCD) using \POWHEG[v2]~\cite{Nason:2009ai,Alioli:2010xd,Nason:2004rx,Frixione:2007vw}, interfaced to \PYTHIA[8]~\cite{Sjostrand:2014zea} for parton showering, hadronisation, and underlying event modelling, with the AZNLO set of tuned parameters~\cite{STDM-2012-23}.
The signal predictions from these MC samples were normalised to the \NNLO (VBF and ZH) or next-to-\NNLO  (ggF) cross-section in QCD plus electroweak corrections at \NLO~\cite{deFlorian:2016spz,Anastasiou:2016cez,Anastasiou:2015ema,Dulat:2018rbf,Harlander:2009mq,Harlander:2009bw,Harlander:2009my,Pak:2009dg,Actis:2008ug,Actis:2008ts,Bonetti:2018ukf}.
The Higgs boson mass was set to \SI{125}{\GeV} for the SM scenario.
For the additional higher-mass Higgs boson scenarios, the mass takes values in the \SIrange{400}{1000}{\GeV} range, in steps of \SI{200}{\GeV}, and in the \SIrange{1000}{3000}{\GeV} range, in steps of \SI{500}{\GeV}.
For the BSM Higgs boson samples, the detector response was simulated using a fast parameterised simulation of the ATLAS calorimeters and the full \GEANT~\cite{Agostinelli:2002hh,SOFT-2010-01} simulation was used for the SM Higgs boson samples.
Additional \pp collisions simulated using \PYTHIA[8]~\cite{Sjostrand:2007gs} with the A3 set of tuned parameters~\cite{ATL-PHYS-PUB-2016-017} were overlaid to simulate the effect of multiple interactions in the same and neighbouring bunch crossings (\pileup).
The simulated events are weighted to reproduce the distribution of the \pileup observed in the data.
These samples are used to check the orthogonality of the channels and produce the final results.
Descriptions of the data and simulation samples used to model backgrounds from SM processes can be found in Refs.~\cite{EXOT-2021-17,HDBS-2019-13,EXOT-2018-63}.
 
\section{Description of the input analyses}
The detailed information about the reconstruction, identification and calibration of physics objects used for analysis selections described below is given in Refs.~\cite{EGAM-2018-01,EGAM-2019-01,EGAM-2021-02,EGAM-2021-01,MUON-2018-03,MUON-2022-01,PERF-2015-09,JETM-2018-05,JETM-2020-03,EXOT-2021-17,HDBS-2019-13,EXOT-2018-63}.
A brief overview of the already published Run~2 searches relevant to the combination performed here is given below.
The \hsmyyd search is performed in the VBF and $ZH$ channels while \hbsmyyd is probed in the VBF and ggF channels.
No significant deviation from the SM prediction is observed in any of these studies.
All searches utilise the signature of an isolated photon and \met, with variations on the number of selected leptons and jets.
 
Events in the VBF channel were selected with the \MET trigger algorithm~\cite{TRIG-2019-01}.
They are further required to have $\met > \SI{150}{\GeV}$ and two VBF jets.
The VBF jets are the leading and subleading transverse momentum (\pt) jets, denoted $j_1$ and $j_2$, satisfying $\pt^{j_1} > \SI{60}{\GeV}$ and $\pt^{j_2} > \SI{50}{\GeV}$, respectively.
In addition, they are required to have a large pseudorapidity separation ($|\Delta\eta_{j_1j_2}| > 3$, $\eta_{j_1}\cdot \eta_{j_2} < 0$), a large invariant mass ($m_{j_1j_2} > \SI{250}{\GeV}$) and not be back-to-back in azimuthal angle ($\Delta\phi_{j_1j_2} < 2$).
An additional jet is allowed with $\pt^{j_3} > \SI{25}{\GeV}$ and a centrality\footnote{
The centrality of an object $i$ in this channel is defined as $
C_i=\text{exp}{\bigg[}-
\frac{4}{(\eta_{j_1}-\eta_{j_2})^2}{\bigg(}\eta_i-\frac{\eta_{j_1}+\eta_{j_2}}{2}{\bigg)}^2{\bigg]}$.
When the object $i$ is centred between the two jets, $C_i$ reaches the maximum value of 1.
When the object $i$ is aligned with one of the two jets, $C_i$ decreases to $1/e$.
} of $C_{j_3} < 0.7$.
Events with a reconstructed lepton ($\ell = e, \mu$) are rejected.
The photon must be isolated and lie in the electromagnetic calorimeter acceptance $|\eta^\gamma| < 2.37$ but not in the transition region ($1.37 < |\eta^\gamma| < 1.52$) between the barrel and endcap.
These isolation and $\eta$ requirements on the photon are used in all three analyses in addition to energy-related criteria specific to the different analyses.
In the VBF analysis discussed here, the photon is required to have $\SI{15}{\GeV} < \phPT < \max(\SI{110}{\GeV}, 0.733 \times \mt)$, where the transverse mass \mt is defined as $\mt(\gamma, \MET) = \sqrt{2\phPT\met \left[1-\cos(\phi_\gamma-\phi_{\met})\right]}$, bounded above by the Higgs boson mass.
The centrality of the photon is required to be $C_{\gamma} > 0.4$.
To increase the sensitivity of the search, events are categorised in ten signal regions (SRs) based on $m_{jj}$ and \mt.
The dominant backgrounds are \Wlvgammajet events, in which the lepton from the $W$ boson decay is lost if it falls outside of the acceptance, and \Zvvgammajet events.
Four dedicated control regions (CRs) are designed to capture the background contributions from \Wevgammajet, \Wmuvgammajet, \Zvvgammajet and processes where electrons are misidentified as fake photons.
Backgrounds due to jets misidentified as fake photons are estimated by using a data-driven method.
Contribution from the ggF process was not considered in the original analysis~\cite{EXOT-2021-17} but is included for this combination.
All SRs and CRs are simultaneously fitted using the profile likelihood fit method to constrain the background contributions and to extract the signal contribution.
 
In the $ZH$ channel, $Z \to \ell^{+}\ell^{-}$ events were selected with single-lepton and dilepton triggers~\cite{TRIG-2018-05,TRIG-2018-01}.
Events are required to contain two same-flavour, oppositely charged (SFOC) leptons with invariant mass $\SI{76}{\GeV} < \mll < \SI{116}{\GeV}$.
Furthermore, they are required to have $\MET > \SI{60}{\GeV}$ and a photon with $\phPT > \SI{25}{\GeV}$ satisfying the isolation and $\eta$ selections.
The selected objects are required to satisfy $\Delta\phi (\met,\pt^{\ell\ell\gamma}) > 2.4$ and $m_{\ell\ell\gamma} > \SI{100}{\GeV}$.
Events with a third lepton with $\pt > \SI{10}{\GeV}$ or any $b$-tagged jet (\SI{77}{\%} working point~\cite{FTAG-2018-01}) are rejected.
To enhance the sensitivity of the search, a boosted decision tree (BDT) algorithm is applied to separate the signal from the backgrounds.
The dominant background contribution comes from \MET mismeasurement in mainly \Zjetgamma and \Zjet processes due to undetected particles or hadronic jets not fully contained in the detector acceptance.
Another major background is from electrons misidentified as photons in the SM diboson processes.
Both of these backgrounds are estimated with data-driven techniques.
The irreducible $VV\gamma$ ($V = W, Z$) background with both $V$ decaying leptonically is constrained in the fit by a dedicated CR\@.
Backgrounds from top quark and Higgs boson production are found to be minor and estimated from MC simulations.
Contributions from ggF and VBF produced Higgs boson processes to the $ZH$ channel are assumed to be negligible and therefore not considered.
A simultaneous fit exploiting the distribution of the BDT discriminant and the single bin $VV\gamma$ CR is performed.
 
\begin{figure}[tb]
\centering
\includegraphics[width=0.5\textwidth]{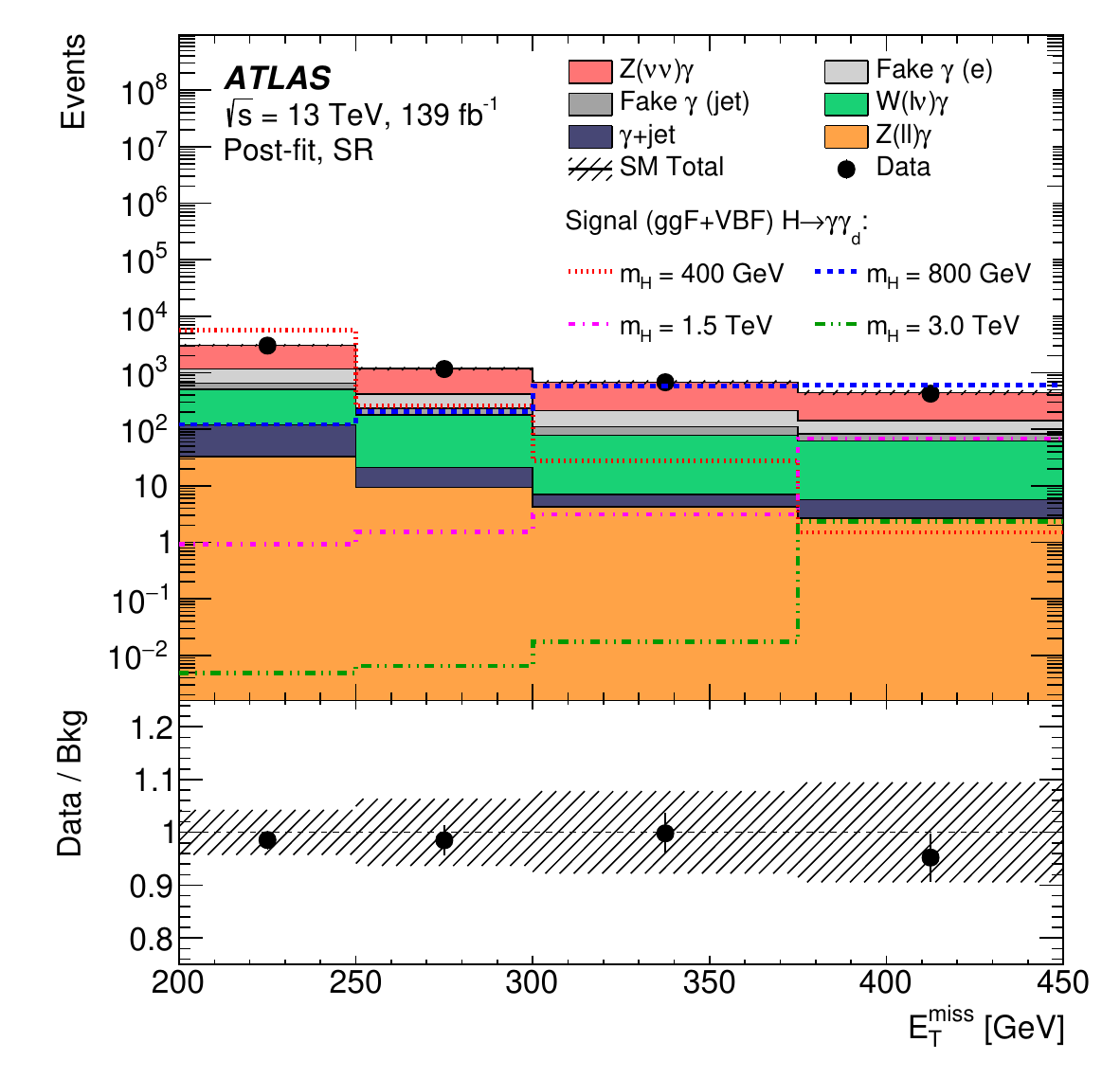}
\caption{
Number of events in each \MET bin in the SR of the ggF channel after a background-only fit.
The signals \hbsmyyd are superimposed onto the observed data and the expected SM background.
The ggF and VBF processes are considered, assuming $\BR(\hbsmyyd) = 5\%$ and the production \xss from Ref.~\cite{deFlorian:2016spz}.
The error bars on data represent the statistical uncertainty, and the shaded band includes statistical and systematic uncertainties determined from the fit.
The lower panel shows the ratio of observed data to expected SM background yields.
The last bin includes the overflow.
This analysis is reinterpreted from the mono-$\gamma$ results in Ref.~\cite{EXOT-2018-63}.
}
\label{fig:ggFdiscriminant}
\end{figure}

Events in the ggF channel were selected with a single photon trigger with a threshold of $\phPT > \SI{140}{\GeV}$~\cite{TRIG-2018-05}.
Events are required to have $\met > \SI{200}{\GeV}$ and at least one photon satisfying the isolation and $\eta$ selections.
The leading photon in the events is required to fulfil the following criteria: the transverse energy $\phPT > \SI{150}{\GeV}$,  the separation between its extrapolated origin and primary vertex of the event along the beam axis $|\Delta z_\gamma| < \SI{250}{\mm}$ and $\Delta\phi(\gamma, \met) > 0.4$.
At most one jet is allowed in the selection if $\pt^{j} > \SI{30}{\GeV}$, $|\eta_{j}| < 4.5$ and $\Delta\phi(j, \met) > 0.4$.
Events that contain leptons ($\ell = e, \mu, \text{hadronically decaying}\ \tau$) are rejected.
To improve the sensitivity, four SRs are defined corresponding to different \MET ranges.
The SM background arises from several processes with either true photons or fake photons from misidentified electrons or jets.
The background with true photons is dominated by the SM \Zvvgamma process.
Secondary contributions come from $\gamma+\text{jets}$ events and from \Wlvgamma and \Zllgamma production with unidentified electrons, muons or with hadronically decaying $\tau$-leptons.
Four CRs are defined to estimate the contributions of these true-photon processes in the SRs through the profile likelihood fit.
The fake photon backgrounds are estimated with control samples that contain electrons or jets, scaled by misidentification rates determined from data.
Signal contributions from both the ggF and VBF Higgs boson processes are considered.
\cref{fig:ggFdiscriminant} shows the \met distribution in the SR after a background-only fit to data.
Signals with higher mass tend to have larger \met.
The highest \met SR captures about \SI{30}{\%} of the events for signal benchmarks with $\mH \ge \SI{1.0}{\TeV}$.

Theoretical uncertainties due to the QCD factorisation and renormalisation scales, and the choice of parton distribution functions are included.
Uncertainties in initial- and final-state radiation due to the choice of parton shower parameters are estimated through simulations with alternative sets of tuned parameters.
 
The event selections for the input channels are summarised in \cref{tab:inputs}.
To perform the combination, a few adjustments are made.
In the VBF channel, the contribution of the ggF process to the BSM Higgs boson decay search is included using the RECAST technique.
The contribution to the signal yields is up to \SI{30}{\%} for the low masses and decreases to \SI{1}{\%} for the high masses.
Additional BSM Higgs boson signal samples are generated to align with the masses probed in the ggF analysis.
The fit regions used in the three analyses are either orthogonal due to variations in the number and type of physics objects required in their final states or have a negligible number of overlapping events due to different jet algorithms being used.
They are therefore treated as statistically independent.
 
\begin{table}[tb]
\centering
\caption{
Summary of the main event selections, discriminating variables and processes considered in the input channels.
The definition of the variables is provided in the text.
}
\label{tab:inputs}
\begin{tabular}{llll}
\toprule
Channels              & VBF       & $ZH$    & ggF  \\
\midrule
Trigger               & \met      & Lepton(s)    & Photon \\
Photons               & $=1$, $C_{\gamma} > 0.4$  & $=1$   & $\ge 1$ \\
$\phPT$ [GeV]         & $\in (15, \max(110, 0.733 \times \mt))$   & $>25$      &  $ >150$ \\
$\met$ [GeV]          & $> 150$   & $> 60$       &  $> 200$  \\
Jets  & 2 or 3, $m_{j_1j_2} > \SI{250}{\GeV}$, $|\Delta\eta_{j_1j_2}| > 3$ & $\le 2$ & $\le 1$ \\
& $\eta_{j_1}\cdot \eta_{j_2} < 0$, $\Delta\phi_{j_1j_2} < 2$, $C_{j_3} < 0.7$  & & \\
Leptons & $=0$ ($e$, $\mu$)      & $=2$, SFOC & $=0$  ($e$, $\mu$, $\tau$)\\
& & $\mll \in (76, 116)~\GeV$ & \\
\midrule
Disc. variables          & $m_{jj}$ and \mt in SR and 4 CRs & BDT score and 1 CR & \met \\
Reference             & \cite{EXOT-2021-17} & \cite{HDBS-2019-13} & \cite{EXOT-2018-63} \\
\midrule
\makecell[l]{Processes considered \\ in the combination}  & VBF, ggF  & $ZH$    & ggF, VBF \\
& & &\\
Combination scenario  & SM, BSM   & SM    & BSM \\
\bottomrule
\end{tabular}
\end{table}

\FloatBarrier
 
\section{Statistical combination}
The results of the combination presented in this paper are obtained from a likelihood function $L(\mu,\vec{\theta})$, where $\mu$ denotes the parameter of interest (POI) of the model, and $\vec{\theta}$ constitutes a set of nuisance parameters, encoding the systematic uncertainty contributions and background normalisation factors that are constrained by CRs in data.
The final likelihood function $L(\mu,\vec{\theta})$ is the product of the likelihoods from individual channels within the combination, which are themselves products of likelihoods computed from the final observables in various categories in a single analysis.
To derive upper limits on the POI, the profile-likelihood-ratio test statistic is used with the \CLs method~\cite{Read:2002hq} following the asymptotic formulae~\cite{Cowan:2010js}.
 
All systematic uncertainties that are considered in the individual analyses are included in the combination and those that stem from common sources are treated as correlated among the input searches.
A complete discussion of their sources can be found in the individual channel publications~\cite{EXOT-2021-17,HDBS-2019-13,EXOT-2018-63}.
Systematic uncertainties related to the data-taking conditions, such as those associated with the integrated luminosity and the modelling of \pileup, are treated as correlated.
Experimental uncertainties related to physics objects used by multiple searches are treated as correlated where appropriate, with the following exceptions.
Uncertainties related to the same object but implemented with different reduced uncertainty schemes are treated as uncorrelated.
In addition, uncertainties that were heavily constrained or pulled in the original input channel are treated as uncorrelated, and the impact on the final upper limits would be $\le \SI{3}{\%}$ if they were correlated.
These constraining powers come from the special phase space probed in certain channels and therefore passing these constraints to the other channels should be avoided.
Uncertainties related to background modelling are considered as uncorrelated since the composition and phase space of the backgrounds are different.
Uncertainties related to signal modelling, stemming from the choice of parton distribution functions and QCD calculations, have minor impact on the final results and are treated as uncorrelated.
Ignoring correlations of the systematic uncertainties between the input channels is found to impact the upper limits by $\le \SI{2}{\%}$.
The \hsmyyd search has similar impacts from the data statistical (\SI{66}{\%}) and systematic (\SI{75}{\%}) uncertainties, where the values are relative to the total uncertainty.
The dominant contributions to the systematic uncertainties stem from the background modelling (\SI{47}{\%}), jet and \met calibration (\SI{40}{\%}), followed by the MC statistical uncertainty (\SI{36}{\%}) and fake-background estimate (\SI{35}{\%}).
All of these impacts are evaluated from the quadratic differences of the fitted errors in the POI before and after fixing the considered nuisance parameters to their best-fit values.
In the \hbsmyyd search, the statistical uncertainty increases from \SI{75}{\%} to \SI{86}{\%} for the higher Higgs boson mass assumptions.
The dominant contributions come from the fake-background estimate, which decreases from \SI{52}{\%} to \SI{29}{\%} depending on the mass, and background modelling uncertainty, ranging from 27\%--38\%.
The uncertainties related to jet and \met, leptons, and MC sample size share a similar $\mysim \SI{20}{\%}$ impact each.
 
\section{Results and interpretation}
\begin{figure}[htb]
\centering
\subfloat[]{\includegraphics[height=15em]{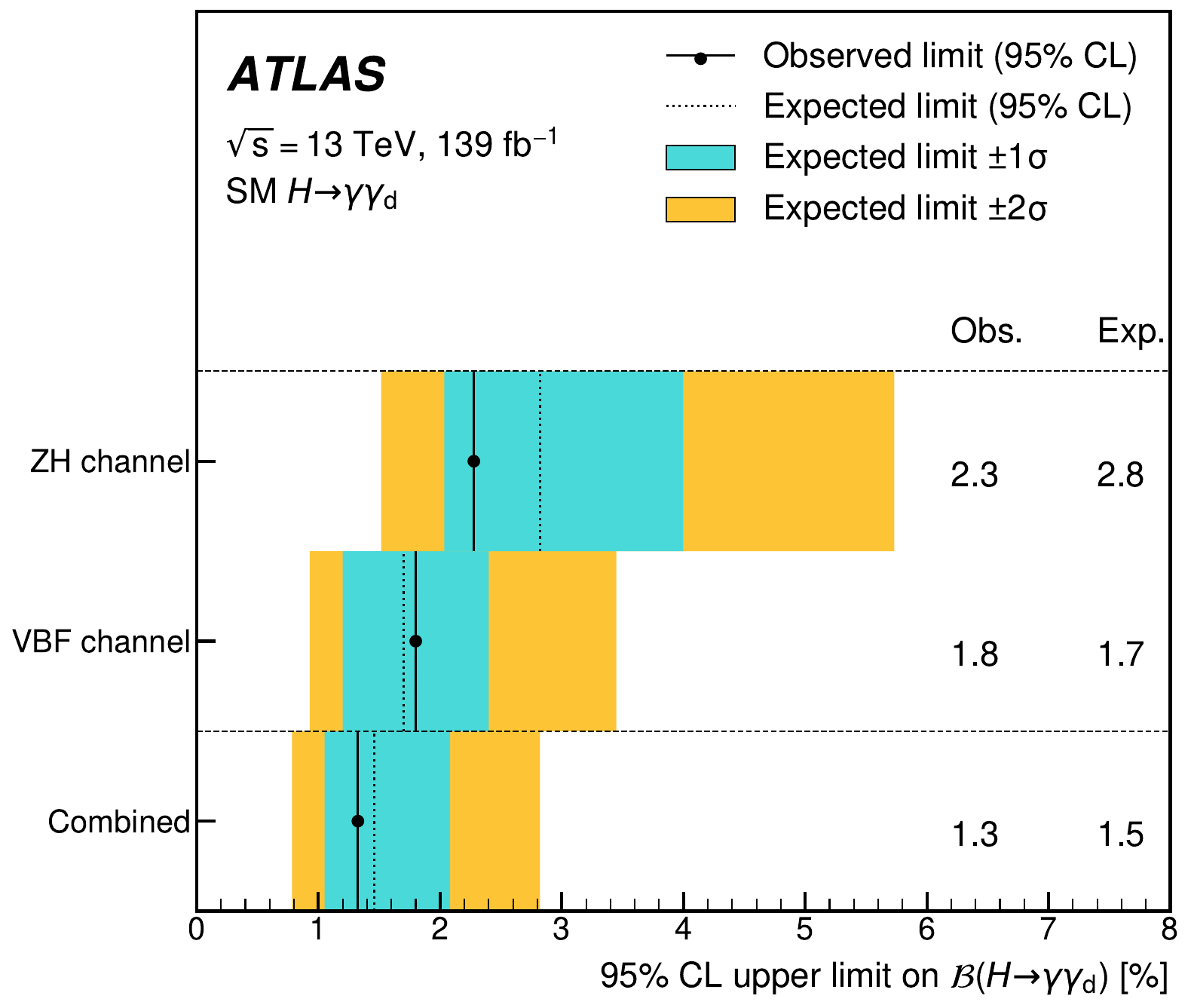}\label{fig:smlimit}} \hspace{2em}
\subfloat[]{\includegraphics[height=15em]{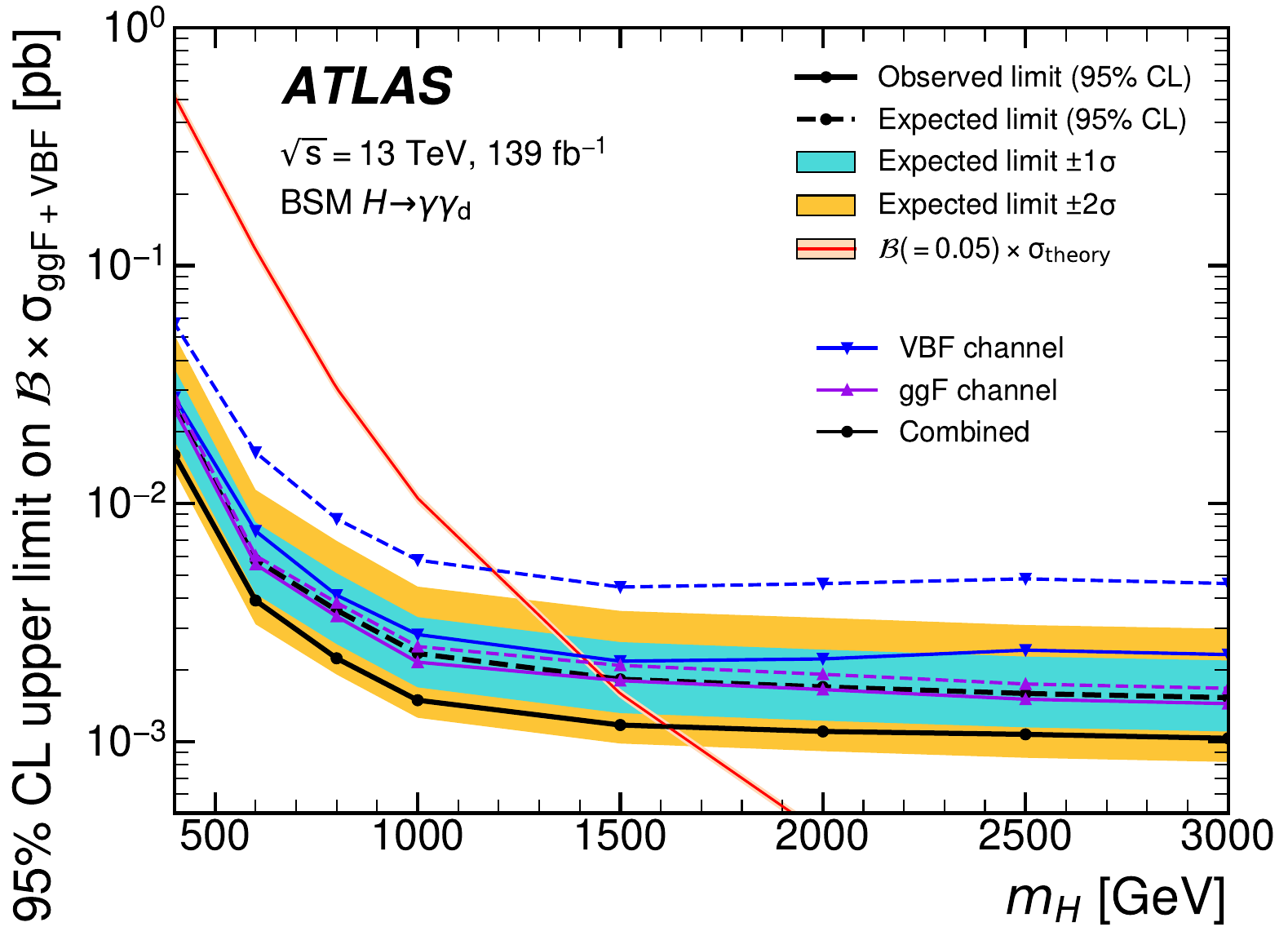}\label{fig:bsmlimit}}
\caption{
Observed and expected 95\% CL upper limits on (a) the branching ratio \BR(\hsmyyd) and (b) \BR(\hbsmyyd) times the BSM Higgs boson ggF + VBF production \xs $\sigma_\text{ggF+VBF}$ as a function of the Higgs boson mass $m_H$.
Results obtained from the input channels are overlaid.
The expected limit and the corresponding error bands are derived assuming the absence of the \hyyd process and with all nuisance parameters profiled to the observed data.
The theory predictions and their error band are taken from Ref.~\cite{deFlorian:2016spz} assuming a branching ratio $\BR(\hbsmyyd) = 5\%$.
}
\end{figure}

This combination yields an observed (expected) 95\% confidence level (CL) upper limit on the branching ratio $\BR(\hsmyyd)$ is $\smbrlimitobs (\smbrlimitexp)$. 
Relative to the most stringent result from the VBF analysis, this combination brings an improvement of \SI{29}{\%} (\SI{14}{\%}) in the sensitivity.
The limits are displayed in \cref{fig:smlimit}.
The observed (expected) 95\% CL upper limits on the BSM Higgs boson production \xs times \BR(\hbsmyyd) as a function of the BSM Higgs boson mass $\mH$, obtained from this combination, are shown in \cref{fig:bsmlimit}.
The observed (expected) limit ranges from \bsmbrlimitobslomass (\bsmbrlimitexplomass) for $\mH = \SI{400}{\GeV}$ to \bsmbrlimitobshimass (\bsmbrlimitexphimass) for $\mH = \SI{3000}{\GeV}$.
Assuming a branching ratio \BR(\hbsmyyd) of $5\%$ and theoretically predicted \xss for \bsmhiggs production from Ref.~\cite{deFlorian:2016spz}, masses of \bsmhiggs below around 1600~\GeV~(1500~\GeV) are excluded.
Relative to the leading ggF channel, the combination with the VBF channel yields an improvement of \SI{33}{\%}~(\SI{14}{\%}) on the \xs times \BR(\hbsmyyd) at $\mH = \SI{1500}{\GeV}$.

Additionally, the results on \BR(\hsmyyd) are interpreted under a minimal simplified model described in Refs.~\cite{Gabrielli:2014oya,Biswas_2022}.
The model assumes the SM prediction for the cross-section and its systematic uncertainties for all Higgs boson production modes. It introduces a generic messenger sector consisting of one left-doublet and one right-singlet of the SU(2)$_\text{L}$ gauge group connecting the dark and the observable sectors.
A minimal scenario can be realised under mass universality in the left and right messenger sector with a mixing parameter $\xi$. Under this scenario, the messenger sector couples to both U(1) and U(1)$_\text{d}$ gauge fields, allowing three Higgs boson decay modes: $\gamma\yd{}$, $\yd\yd{}$ and $\gamma\gamma$.
Their branching ratios are all interconnected and can be expressed using two free, dimensionless parameters: $\xi$ and the fine structure constant \alphad associated with the U(1)$_\text{d}$ gauge group~\cite{Biswas_2022}, whose values are zero in the SM.
Hence, setting upper limits on, or measuring the value of the branching ratio of the Higgs boson decaying into $\gamma\gamma$, $\gamma\yd$, and $\yd\yd$ can be translated into restrictions on the allowed parameter space in the \plane.
The range $0 < \alphad < 1$ and $0 \leq \xi < 1$ is explored in this study.
An extra parameter $\chi = \pm 1$ accounts for the relative sign of the new physics and SM contributions in the amplitudes of the $\smhiggs\to\gamma\gamma$ and $\smhiggs\to gg$ decays.
It provides two scenarios with either constructive ($\chi=+1$) or destructive ($\chi=-1$) interference between the messenger sector and the SM particles affecting the \BR($\smhiggs\to\gamma\gamma$).
 
\begin{figure}[htb]
\centering
\includegraphics[width=0.7\textwidth]{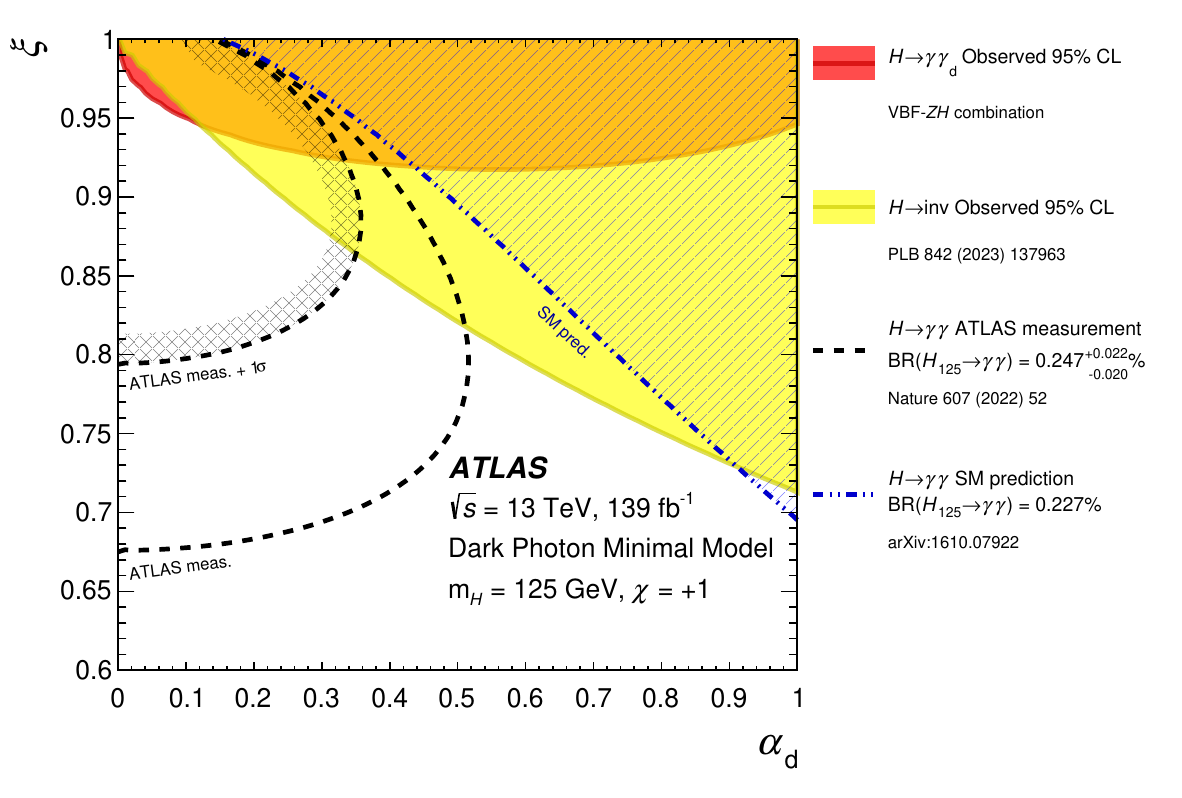}
\caption{
Observed 95\% CL exclusion regions in the $(\alphad,\xi)$ parameter space from the VBF-$ZH$  combination of $\smhiggs \to \gamma\yd$ searches (dark coloured area) and the reinterpretation of the ATLAS measured limit of \BR($\smhiggs\to \mathrm{invisible}$) as $\smhiggs \to \yd\yd$ (light coloured area).
Both the central value (dashed line) and the $+1\sigma$ upper bound (dashed-hashed line) show the less favoured \plane{} obtained from the current ATLAS measurement of $\BR(\smhiggs \to \gamma\gamma)$. The dash-dotted line and the associated dashed area show the parameter space that is not allowed by the SM \BR$(\smhiggs \to \gamma\gamma)$ predictions assuming positive interference with the minimal messenger sector. The parameter $\chi = +1$ parameterises the relative sign of the new physics and SM contributions in the amplitudes of the $\smhiggs \to\gamma\gamma$ and $\smhiggs \to gg$ decays.
}
\label{fig:interpretation}
\end{figure}

\Cref{fig:interpretation} shows the observed exclusion limits derived from the combination of the VBF and $ZH$ channels for \hsmyyd and from the reinterpretation of the ATLAS search for $\smhiggs \to\text{invisible}$~\cite{HIGG-2021-05}.
In addition, the ATLAS measurement of $\BR(\smhiggs \to \gamma\gamma) = 0.247^{+0.022}_{-0.020} \%$~\cite{HIGG-2021-23} provides a one standard deviation ($\sigma$) upper bound. The SM predicted $\BR(\smhiggs \to \gamma\gamma)$ defines a region in the $(\alphad, \xi$) parameter space of the minimal model that is not allowed when assuming a positive interference from the messenger sector in the $\gamma\gamma$ final state ($\chi=+1$).
 
From the reinterpretation of the $\BR(\smhiggs \to\text{invisible})$ result in terms of a $\smhiggs \to \yd\yd$ signal, values down to about $\xi \simeq 0.7$ at $\alphad = 1$ are excluded for $\chi=+1$. The $\smhiggs \to \gamma\yd$ combination presented here provides additional sensitivity in the low \alphad region, which is also disfavoured by the $\BR(\smhiggs \to \gamma\gamma)$ measurement. The $\chi=-1$ scenario has also been investigated but no constraints can be placed in this case.
 
\FloatBarrier
\section{Conclusion}
In conclusion, this paper reports a combined search for \hyyd produced in ggF, VBF, and $ZH$ processes using \SI{139}{\ifb} of \pp collision data at $\sqrt{s} = \SI{13}{\TeV}$ collected by the ATLAS detector at the LHC\@.
Various Higgs boson mass assumptions are probed including the SM value of \SI{125}{\GeV} and higher values ranging from \SI{400}{\GeV} to \SI{3}{\TeV}.
The combined analysis sets an observed (expected) 95\% CL upper limit on \BR(\hsmyyd) at 1.3 (1.5)\,\%.
The observed (expected) 95\% confidence level limit on the \xs times branching ratio ranges from \bsmbrlimitobslomass (\bsmbrlimitexplomass) for $\mH = \SI{400}{\GeV}$ to \bsmbrlimitobshimass (\bsmbrlimitexphimass) for $\mH = \SI{3}{\TeV}$.
This study provides the most stringent constraints on Higgs bosons decaying into a photon and a massless dark photon to date.
The results are also interpreted in the context of a minimal simplified model, providing complementary exclusion in the parameter space obtained from other searches for this model.


\FloatBarrier
 
\section*{Acknowledgements}


We thank CERN for the very successful operation of the LHC and its injectors, as well as the support staff at
CERN and at our institutions worldwide without whom ATLAS could not be operated efficiently.
 
The crucial computing support from all WLCG partners is acknowledged gratefully, in particular from CERN, the ATLAS Tier-1 facilities at TRIUMF/SFU (Canada), NDGF (Denmark, Norway, Sweden), CC-IN2P3 (France), KIT/GridKA (Germany), INFN-CNAF (Italy), NL-T1 (Netherlands), PIC (Spain), RAL (UK) and BNL (USA), the Tier-2 facilities worldwide and large non-WLCG resource providers. Major contributors of computing resources are listed in Ref.~\cite{ATL-SOFT-PUB-2023-001}.
 
We gratefully acknowledge the support of ANPCyT, Argentina; YerPhI, Armenia; ARC, Australia; BMWFW and FWF, Austria; ANAS, Azerbaijan; CNPq and FAPESP, Brazil; NSERC, NRC and CFI, Canada; CERN; ANID, Chile; CAS, MOST and NSFC, China; Minciencias, Colombia; MEYS CR, Czech Republic; DNRF and DNSRC, Denmark; IN2P3-CNRS and CEA-DRF/IRFU, France; SRNSFG, Georgia; BMBF, HGF and MPG, Germany; GSRI, Greece; RGC and Hong Kong SAR, China; ISF and Benoziyo Center, Israel; INFN, Italy; MEXT and JSPS, Japan; CNRST, Morocco; NWO, Netherlands; RCN, Norway; MNiSW, Poland; FCT, Portugal; MNE/IFA, Romania; MESTD, Serbia; MSSR, Slovakia; ARRS and MIZ\v{S}, Slovenia; DSI/NRF, South Africa; MICINN, Spain; SRC and Wallenberg Foundation, Sweden; SERI, SNSF and Cantons of Bern and Geneva, Switzerland; MOST, Taipei; TENMAK, T\"urkiye; STFC, United Kingdom; DOE and NSF, United States of America.
 
Individual groups and members have received support from BCKDF, CANARIE, CRC and DRAC, Canada; CERN-CZ, FORTE and PRIMUS, Czech Republic; COST, ERC, ERDF, Horizon 2020, ICSC-NextGenerationEU and Marie Sk{\l}odowska-Curie Actions, European Union; Investissements d'Avenir Labex, Investissements d'Avenir Idex and ANR, France; DFG and AvH Foundation, Germany; Herakleitos, Thales and Aristeia programmes co-financed by EU-ESF and the Greek NSRF, Greece; BSF-NSF and MINERVA, Israel; NCN and NAWA, Poland; La Caixa Banking Foundation, CERCA Programme Generalitat de Catalunya and PROMETEO and GenT Programmes Generalitat Valenciana, Spain; G\"{o}ran Gustafssons Stiftelse, Sweden; The Royal Society and Leverhulme Trust, United Kingdom.
 
In addition, individual members wish to acknowledge support from Armenia: Yerevan Physics Institute (FAPERJ); CERN: European Organization for Nuclear Research (CERN PJAS); Chile: Agencia Nacional de Investigaci\'on y Desarrollo (FONDECYT 1230812, FONDECYT 1230987, FONDECYT 1240864); China: Chinese Ministry of Science and Technology (MOST-2023YFA1605700), National Natural Science Foundation of China (NSFC - 12175119, NSFC 12275265, NSFC-12075060); Czech Republic: Czech Science Foundation (GACR - 24-11373S), Ministry of Education Youth and Sports (FORTE CZ.02.01.01/00/22\_008/0004632), PRIMUS Research Programme (PRIMUS/21/SCI/017); EU: H2020 European Research Council (ERC - 101002463); European Union: European Research Council (ERC - 948254, ERC 101089007), Horizon 2020 Framework Programme (MUCCA - CHIST-ERA-19-XAI-00), European Union, Future Artificial Intelligence Research (FAIR-NextGenerationEU PE00000013), Italian Center for High Performance Computing, Big Data and Quantum Computing (ICSC, NextGenerationEU); France: Agence Nationale de la Recherche (ANR-20-CE31-0013, ANR-21-CE31-0013, ANR-21-CE31-0022), Investissements d'Avenir Labex (ANR-11-LABX-0012); Germany: Baden-Württemberg Stiftung (BW Stiftung-Postdoc Eliteprogramme), Deutsche Forschungsgemeinschaft (DFG - 469666862, DFG - CR 312/5-2); Italy: Istituto Nazionale di Fisica Nucleare (ICSC, NextGenerationEU); Japan: Japan Society for the Promotion of Science (JSPS KAKENHI JP22H01227, JSPS KAKENHI JP22H04944, JSPS KAKENHI JP22KK0227, JSPS KAKENHI JP23KK0245); Netherlands: Netherlands Organisation for Scientific Research (NWO Veni 2020 - VI.Veni.202.179); Norway: Research Council of Norway (RCN-314472); Poland: Polish National Agency for Academic Exchange (PPN/PPO/2020/1/00002/U/00001), Polish National Science Centre (NCN 2021/42/E/ST2/00350, NCN OPUS nr 2022/47/B/ST2/03059, NCN UMO-2019/34/E/ST2/00393, UMO-2020/37/B/ST2/01043, UMO-2021/40/C/ST2/00187, UMO-2022/47/O/ST2/00148, UMO-2023/49/B/ST2/04085); Slovenia: Slovenian Research Agency (ARIS grant J1-3010); Spain: Generalitat Valenciana (Artemisa, FEDER, IDIFEDER/2018/048), Ministry of Science and Innovation (MCIN \& NextGenEU PCI2022-135018-2, MICIN \& FEDER PID2021-125273NB, RYC2019-028510-I, RYC2020-030254-I, RYC2021-031273-I, RYC2022-038164-I), PROMETEO and GenT Programmes Generalitat Valenciana (CIDEGENT/2019/027); Sweden: Swedish Research Council (Swedish Research Council 2023-04654, VR 2018-00482, VR 2022-03845, VR 2022-04683, VR 2023-03403, VR grant 2021-03651), Knut and Alice Wallenberg Foundation (KAW 2018.0157, KAW 2018.0458, KAW 2019.0447, KAW 2022.0358); Switzerland: Swiss National Science Foundation (SNSF - PCEFP2\_194658); United Kingdom: Leverhulme Trust (Leverhulme Trust RPG-2020-004), Royal Society (NIF-R1-231091); United States of America: U.S. Department of Energy (ECA DE-AC02-76SF00515), Neubauer Family Foundation.


\clearpage
-------------------------------------------------------------------------------

\printbibliography
 
\clearpage
 
\begin{flushleft}
\hypersetup{urlcolor=black}
{\Large The ATLAS Collaboration}

\bigskip

\AtlasOrcid[0000-0002-6665-4934]{G.~Aad}$^\textrm{\scriptsize 104}$,
\AtlasOrcid[0000-0001-7616-1554]{E.~Aakvaag}$^\textrm{\scriptsize 17}$,
\AtlasOrcid[0000-0002-5888-2734]{B.~Abbott}$^\textrm{\scriptsize 123}$,
\AtlasOrcid[0000-0002-0287-5869]{S.~Abdelhameed}$^\textrm{\scriptsize 119a}$,
\AtlasOrcid[0000-0002-1002-1652]{K.~Abeling}$^\textrm{\scriptsize 56}$,
\AtlasOrcid[0000-0001-5763-2760]{N.J.~Abicht}$^\textrm{\scriptsize 50}$,
\AtlasOrcid[0000-0002-8496-9294]{S.H.~Abidi}$^\textrm{\scriptsize 30}$,
\AtlasOrcid[0009-0003-6578-220X]{M.~Aboelela}$^\textrm{\scriptsize 45}$,
\AtlasOrcid[0000-0002-9987-2292]{A.~Aboulhorma}$^\textrm{\scriptsize 36e}$,
\AtlasOrcid[0000-0001-5329-6640]{H.~Abramowicz}$^\textrm{\scriptsize 154}$
% ...
\bigskip
\\

% ...
$^{17}$Department for Physics and Technology, University of Bergen, Bergen; Norway.\\
% ...
$^{30}$Physics Department, Brookhaven National Laboratory, Upton NY; United States of America.\\
% ...
$^{36}$$^{(a)}$Facult\'e des Sciences Ain Chock, R\'eseau Universitaire de Physique des Hautes Energies - Universit\'e Hassan II, Casablanca;$^{(b)}$Facult\'{e} des Sciences, Universit\'{e} Ibn-Tofail, K\'{e}nitra;$^{(c)}$Facult\'e des Sciences Semlalia, Universit\'e Cadi Ayyad, LPHEA-Marrakech;$^{(d)}$LPMR, Facult\'e des Sciences, Universit\'e Mohamed Premier, Oujda;$^{(e)}$Facult\'e des sciences, Universit\'e Mohammed V, Rabat;$^{(f)}$Institute of Applied Physics, Mohammed VI Polytechnic University, Ben Guerir; Morocco.\\
% ...
$^{45}$Physics Department, Southern Methodist University, Dallas TX; United States of America.\\
% ...
$^{50}$Fakult\"{a}t Physik , Technische Universit{\"a}t Dortmund, Dortmund; Germany.\\
% ...
$^{56}$II. Physikalisches Institut, Georg-August-Universit\"{a}t G\"ottingen, G\"ottingen; Germany.\\
% ...
$^{104}$CPPM, Aix-Marseille Universit\'e, CNRS/IN2P3, Marseille; France.\\
% ...
$^{119}$$^{(a)}$New York University Abu Dhabi, Abu Dhabi;$^{(b)}$United Arab Emirates University, Al Ain; United Arab Emirates.\\
% ...
$^{123}$Homer L. Dodge Department of Physics and Astronomy, University of Oklahoma, Norman OK; United States of America.\\
% ...
$^{154}$Raymond and Beverly Sackler School of Physics and Astronomy, Tel Aviv University, Tel Aviv; Israel.\\
% ...
$^{a}$ Also Affiliated with an institute covered by a cooperation agreement with CERN.\\
$^{b}$ Also at An-Najah National University, Nablus; Palestine.\\
$^{c}$ Also at Borough of Manhattan Community College, City University of New York, New York NY; United States of America.\\
$^{d}$ Also at Center for Interdisciplinary Research and Innovation (CIRI-AUTH), Thessaloniki; Greece.\\
$^{e}$ Also at CERN, Geneva; Switzerland.\\
$^{f}$ Also at CMD-AC UNEC Research Center, Azerbaijan State University of Economics (UNEC); Azerbaijan.\\
% ...

\end{flushleft}


\end{document}